\documentclass[12pt]{revtex4}
  
\usepackage{hyperref}
\usepackage{amsmath,amsfonts,amssymb}
\hypersetup{dvips,dvipdfm,colorlinks=true,urlcolor=magenta,filecolor=magenta,linktoc=page,citecolor=red,linkcolor=blue,bookmarks=true}
\usepackage{graphicx,epstopdf}

\begin{document}

\title{Does fractal Universe describe a complete cosmic scenario $?$}

\author{Dipanjana Das$^1$\footnote {ddipanjana91@gmail.com}}
\author{Sourav Dutta$^2$\footnote {sduttaju@gmail.com}}
\author{Abdulla Al Mamon$^1$\footnote {abdulla.physics@gmail.com}}
\author{Subenoy Chakraborty$^1$\footnote {schakraborty.math@gmail.com}}
\affiliation{$^1$Department of Mathematics, Jadavpur University, Kolkata-700032, West Bengal, India\\
	$^2$ Department of Pure Mathematics, Ballygunge Science College, 35, Ballygunge Circular Rd, Ballygunge, Kolkata, West Bengal 700019}


\begin{abstract}
The present work deals with evolution of the fractal model of the Universe in the background of homogeneous and isotropic FLRW space--time geometry. The cosmic substrum is taken as perfect fluid with barotropic equation of state. A general prescription for the deceleration parameter is determined and it is examined whether the deceleration parameter may have more than one transition during the evolution of the fractal Universe for monomial form of the fractal function as a function of the scale factor. Finally, the model has been examined by making comparison with the observed data.	\\\\

Keywords: Fractal Universe, Cosmic Scenario, Deceleration Parameter.

\end{abstract}
\maketitle

\section{introduction}

The homogeneous and isotropic FLRW (Friedmann Lemaitre Robertson Walker) model without the Lambda term in standard relativistic cosmology has been put in a big question mark for the last two decades due to a  series of observational evidences (\cite{Ag}, \cite{sp}, \cite{wj}, \cite{pa}, \cite{Ag1}, \cite{mg}, \cite{us}, \cite{dn}, \cite{acs}, \cite{jh}, \cite{ec}, \cite{dj}, \cite{bj}, \cite{vs}). However, there are various attempts to go beyond the simple hypothesis of homogeneity and isotropy and to include the effect of spatial inhomogeneities in the metric. These observational data are interpreted in the framework of the Friedmann solutions of the Einstein field equations and it is found that there should be an acceleration. This implies the presence of a Lambda--like term in the Friedmann equations or alternatively, this could be due to the effect of large scale spatial inhomogeneities (\cite{dj}). Thus this challenge of cosmology has been addressed in the frame work of Einstein gravity by introducing inhomogeneous space time model (\cite{tbuchert}, \cite{tbuchert1}, \cite{tbuchert2}) or by modifying the Einstein gravity itself in the usual homogeneous and isotropic FLRW model. The modifications in the Einstein field equations (for FLRW model) has been done either by introducing (in the right hand side) some exotic matter (\cite{vs}, \cite{tp}) (known as DE) having large negative pressure or by modifying the geometric part of the field equations i.e., considering modified gravity theories (considering general form of the Lagrangian density than Einstein--Hilbert or by higher dimensional theories) (\cite{tclifton}).  To have an idea of what the ``fractal effects" could possible be, an attempt is made in the present work to explain the present (observed) cosmic acceleration in a gravity theory with imprints of fractal effects. \\

One of the challenging issues in the present day theoretical physics is to formulate a consistent theory of quantum gravity. Besides the major attempts namely string theory and (loop) quantum gravity, other independent investigations such as causal dynamical triangulations (\cite{jambjorn1}), asymptotically safe gravity (\cite{olauscher}), spin--foam models (\cite{lmodesto}, \cite{dbenedetti}) and Ho\v{r}ava--Lifshitz gravity (\cite{phorava}, \cite{phorava1}) have received some attention. All these theories exhibit a running spectral dimension $d_s$ (\cite{salex}, \cite{rrammal}, \cite{jambjorn}, \cite{dben}) of spacetime such that $d_s$ is smaller than four in the UV scale (\cite{phorava1}, \cite{gcal}). Systems whose effective dimensionality changes with the scale may have fractal behaviour, even if they are defined on a smooth manifold. A lower spectral dimension can be associated with scenarios with improved renormalization but it is not true in general.\\
In fractal Universe time and space--coordinates scale isotropically  i.e., $[x^{\mu}]=-1,~\mu=0, 1, ......\mbox{D}-1$ and the standard measure in the action is replaced by a non--trivial measure (usually appears in Lebesgue--Stieltjes integrals):

$$d^D x \rightarrow d \rho(x),~[\rho]=-D\alpha\neq -D,$$
where $D$ is the topological (positive integer) dimension of the embedding space--time and the parameter $\alpha> 0$ roughly corresponds to the fraction of states preserved at a given time during the evolution of the system. One can obtain such structure through variable effective dimensionality of the Universe at different scales. This feature can be obtained simply by introducing fractal action as on net fractals and they can be approximated by fractal integrals (\cite{gcal}). Usually, the measure isx considered on a fractal set as general Borel probability measure $\rho$. So in $D$ dimensions, $(M, \rho)$ denotes the metric space--time $M$ equipped with measure $\rho$. Here $\rho$ is absolutely continuous with 
$$d\rho=\left( d^D x\right)v(x),$$
some multidimensional Lebesgue measure and $v$ is the weight function (fractal function).   \\

In a fractal space--time the total action of Einstein gravity can be written as 

\begin{equation}
S=S_g+S_m, \label{f1}
\end{equation}
where 

\begin{equation} 
S_g=\frac{M_p^2}{2}\int d^D x ~v(x) \sqrt{-g}\left(R-2 \Lambda-w \partial_{\mu} v \partial^{\mu} v\right), \label{f2}
\end{equation}
is the gravitational part of the action \cite{gcal}, \cite{gcal1} and 

\begin{equation}
S_m=\int d^D x ~v(x) \sqrt{-g} \mathcal{L}_m, \label{f3}
\end{equation}
is the action of the matter part minimally coupled to gravity. Here $M_p=(8 \pi G)^{-\frac{1}{2}}$ is the reduced Planck mass and $w$  stands for fractal parameter in the action in (i.e., equation (\ref{f2})). The paper is organized as follows: section II presents fractal Universe as standard cosmology with interacting two fluids. In section III, fractal cosmology has been shown to be equivalent as a particle creation mechanism in the context of non equilibrium thermodynamical prescription and temperature of the individual fluids has been determined in terms of creation rate parameter. Also entropy variation of the individual fluids as well as the total entropy variation has been evaluated for this open thermodynamical system. A detailed study of the evolution of the deceleration parameter and its possible ranges in different stage of evolution has been studied in section IV. In section V, the model has been compared with observational data using contour plots and variation of the deceleration parameter has been shown graphically. The paper ends with a brief summary in section VI.

\section {Fractal Universe as interacting two fluids in Einstein gravity}
In homogeneous, isotropic and flat FLRW space--time model the variation of the action (in equation (\ref{f1})) with respect to the metric tensor gives the Friedmann equations in a fractal Universe as 

\begin{equation}
3H^2=8\pi G\rho-3H \frac{\dot{v}}{v}+\frac{w}{2}\dot{v}^2+\Lambda,\label{f4}
\end{equation}
and
\begin{equation}
6(\dot{H}+H^2)=-8 \pi G (\rho+3p)+6H \frac{\dot{v}}{v}-2w \dot{v}^2+3\frac{\Box v}{v}+2\Lambda,\label{f5}
\end{equation}
where $`\Box$' is the usual D'Alembertian operator and matter is chosen as perfect fluid with barotropic equation of state: $p= \gamma \rho$. For simplicity we choose $\Lambda=0$ and units are chosen such that $8\pi G=1.$ The above field equations can be written as (\cite{sh}) 

\begin{equation}
3H^2=\rho+\rho_f=\rho_t,\label{f4.1}
\end{equation}
and
\begin{equation}
2\dot{H}+3H^2=-p-p_f=-p_t,\label{f5.1}
\end{equation}
where $(\rho_f, p_f)$ are termed as energy density and thermodynamic pressure of the effective (hypothetical) fractal fluid and have the expressions

\begin{equation}
\rho_f=\frac{w}{2}\dot{v}^2-3H\frac{\dot{v}}{v},\label{f6}
\end{equation}
and
\begin{equation}
p_f=\frac{w}{2}\dot{v}^2+2H\frac{\dot{v}}{v}+\frac{\ddot{v}}{v}.\label{f7}
\end{equation}

The energy conservation equation in a fractal Universe takes the form (\cite{gcal}, \cite{sh}, \cite{ash})

\begin{equation}
\dot{\rho}+\left(3H+\frac{\dot{v}}{v}\right)\left(\rho+p\right)=0.\label{f8}
\end{equation}

Note that if the fractal function $v$ is chosen as constant then from equations (\ref{f6}) and (\ref{f7}) $\rho_f=0=p_f$ and equations (\ref{f4.1}), (\ref{f5.1}) and (\ref{f8}) represent the usual Einstein field equations and the energy conservation law for Einstein gravity in FLRW model.\\

From the field equations (\ref{f4.1}) and (\ref{f5.1}), due to Bianchi identity we have 

\begin{equation}
\dot{\rho_t}+3H(\rho_t+p_t)=0.\label{f9}
\end{equation}

Now from equations (\ref{f8}) and (\ref{f9}) one can write the individual matter conservation equations as

\begin{equation}
\dot{\rho}+3H(\rho+p)=-\frac{\dot{v}}{v}(\rho+p)=Q,\label{f10}
\end{equation}
and
\begin{equation}
\dot{\rho_f}+3H(\rho_f+p_f)=-Q.\label{f11}
\end{equation}

Thus, the modified Friedmann equations in fractal Universe can be interpreted as Friedmann equations in Einstein gravity for an interacting two fluid system of which one is the usual normal fluid under consideration and other is the effective fractal fluid and the interaction term is given by $Q=-\frac{\dot{v}}{v}(\rho+p).$\\

However, one can write the above conservation equations (\ref{f10}) and (\ref{f11}) in non--interacting form with effective equation of state parameters as 

\begin{equation}
\gamma^{(eff)}=\gamma+\frac{\dot{v}}{3Hv}(1+\gamma),\label{f12}
\end{equation}
\begin{equation} 
\gamma_f^{(eff)}=\gamma_f-\frac{\dot{v}(1+\gamma)}{3Hv}r,\label{f13}
\end{equation}
where $r=\frac{\rho}{\rho_f}$ is the dimensionless coincidence parameter and $\gamma_f=\frac{p_f}{\rho_f}$  is the equation of state parameter for effective fractal fluid.

Usually, for interacting two fluid system the interaction term $Q$ should be positive as energy is transferred to the dark matter (the usual fluid here). Here positivity of $Q$ implies that the fractal function $v$ should decrease with the evolution ($i.e., \dot{v}<0$). Thus, if the present normal fluid is chosen as cold dark matter ($i.e, \gamma=0$) then $\gamma^{(eff)}<0$ $i.e.,$ we have effectively some exotic nature of the matter component due to fractal cosmology. However, if the fluid is chosen as hot dark matter ($i.e., \gamma>0$) then it  may or may not remain hot dark matter in fractal cosmology, depending on the explicit choice of the fractal function.\\

Moreover, using the conservation equations (\ref{f10}) and (\ref{f11}) the time evolution of the coincidence parameter may be written as 

\begin{equation} 
\frac{dr}{d \tau}=r\left[(\gamma_f-\gamma)-\frac{1+\gamma}{v}(1+r)\frac{dv}{d\tau}\right],\label{f14}
\end{equation}
with $\tau=3 \ln a$. As a result the time variation of the total energy density can be written as 

\begin{equation} 
\frac{d \rho_T}{d \tau}=-\rho_T\left[1+\frac{r\gamma+\gamma_f}{1+r}\right].\label{f15}
\end{equation}

\section{Fractal Universe as particle creation mechanism and temperature of the fluid components}
The Friedmann equations (\ref{f4.1}) and (\ref{f5.1}) of the last section for the fractal Universe can be written as
\begin{eqnarray} 
3H^2&=&\rho+\rho_f=-\rho_t,\nonumber\\
2\dot{H}+3H^2&=&-(p+p_c)-(p_f+p_{cf}),\label{f16}
\end{eqnarray}
with $p_{cf}=-p_c$, the dissipative pressure (bulk viscous pressure) for the fluid components. So the energy conservation equations (\ref{f10}) and (\ref{f11}) can be written as 
 \begin{eqnarray} 
 \dot{\rho}+3H(\rho+p+p_c)&=&0,\nonumber\\ 
 \dot{\rho_f}+3H(\rho_f+p_f+p_{cf})&=&0,\label{f17}
 \end{eqnarray}
 with
 \begin{equation} 
 p_c=\frac{\dot{v}}{3Hv}(\rho+p)=-p_{cf}.\label{f18}
 \end{equation}
 
 In the context of non--equilibrium thermodynamics this dissipative pressure may be assumed to be due to particle creation process. So the particle number conservation relations take the form
 
 \begin{equation} 
 \dot{n}+3Hn=\Gamma_n,\label{f19}
 \end{equation}
 and
 \begin{equation} 
 \dot{n_f}+3Hn_f=-\Gamma_f n_f.\label{f20}
 \end{equation}
 
Here `$n$' represent the normal fluid particle density with `$n_f$' corresponds to number density of the effective particles (termed as fractal particle) for the effective fluid. Note that these effective particles are introduced to map the present model to one for a standard GR Universe with some particle species. It is assumed that normal fluid particles are created (i.e., the particle creation rate $ \Gamma>0$) and the effective `fractal' particles are annihilated (i.e., $\Gamma_f<0$) in course of the evolution. For simplicity if we assume the non--equilibrium thermodynamical process to be adiabatic (i.e., isentropic) then the dissipative pressures are related to the particle creation rates linearly as (\cite{ssaha}) 
 
 \begin{equation} 
 p_c=-\frac{\Gamma}{3H}(\rho+p)~~~\mbox{and}~~~p_{cf}=\frac{\Gamma_{f}}{3H}(\rho_f+p_f).\label{f21}
 \end{equation}
 
 Now comparing equations  (\ref{f18}) and (\ref{f21}) the particle creation rates are given by
 
 \begin{equation} 
 \Gamma=\frac{\dot{v}}{v}~~~\mbox{and}~~~\Gamma_f=-\frac{\dot{v} (\rho+p)}{v (\rho_f+p_f)}.\label{f22}
 \end{equation}
 
 Thus, in fractal Universe the particle creation rate is related to the fractal function by the above relations.\\
 
 For the combined single fluid as $p_{ct}=p_c+p_{cf}=0$ so the particle creation rate for the resulting single fluid should vanish identically. Hence effective single fluid forms a closed system.\\
 
 Further, due to particle creation mechanism there is an energy transfer between the two fluid components and as a result, the two subsystems may have different temperatures and thermodynamics of irreversible process comes into the picture.\\
 
 Using Euler's thermodynamical relation: $n T s=\rho+p,$ the evolution of the temperature of the individual fluid is given by (\cite{sc}, \cite{jz})
 
 \begin{equation} 
 \frac{\dot{ T}}{T}=-3H\left(\gamma^{(eff)}+\frac{\Gamma}{3H}\right)+\frac{\dot{\gamma}}{(1+\gamma)},\label{f23}
 \end{equation}
 and
 \begin{equation} 
 \frac{\dot{ T_f}}{T_f}=-3H\left(\gamma_f^{(eff)}-\frac{\Gamma_f}{3H}\right)+\frac{\dot{\gamma_f}}{(1+\gamma_f)},\label{f24}
 \end{equation}
 where $\gamma^{(eff)}$ and $\gamma_f^{(eff)}$ defined in equations (\ref{f12}) and (\ref{f13}) can now be written in terms of particle creation rate as 
 
  \begin{equation} 
 \gamma^{(eff)}=\gamma-\frac{\Gamma}{3H}(1+\gamma),~\gamma_f^{(eff)}=\gamma_f+\frac{\Gamma_f}{3H}(1+\gamma_f) .\label{f25}
  \end{equation}
 
 Now integrating equations (\ref{f23}) and (\ref{f24}) the temperature of the individual fluid component can be written as (\cite{sc})
 
  \begin{eqnarray} 
 T&=&T_0(1+\gamma) exp\left[-3 \int_{a_0}^a \gamma \left(1-\frac{\Gamma}{3H}\right)\frac{da}{a}\right],\nonumber\\
T_f&=&T_0(1+\gamma_f) exp\left[-3 \int_{a_0}^a \gamma_f \left(1+\frac{\Gamma_f}{3H}\right)\frac{da}{a}\right], \label{f26}
  \end{eqnarray}
 where $T_0$ is the common temperature of the two fluids in equilibrium configuration and $a_0$ is the value of the scale factor in the equilibrium state. In particular, using equation (\ref{f22}) for $\Gamma$ (in terms of the fractal function) the temperature of the normal fluid for constant $\gamma$ can have the following explicit expression 
 \begin{equation} 
T=T_0(1+\gamma)\left(\frac{a}{a_0}\right)^{-3\gamma} \left(\frac{v}{v_0}\right)^{-\gamma}.\label{f27}
 \end{equation}
 
 Usually, at very early stages of the evolution of the Universe one should have $T_f<T$ and then with the evolution of the Universe, both the cosmic fluids attain an equilibrium era at $a=a_0$ with $T=T_f=T_0$. In the subsequent evolution of the Universe one has $a>a_0$ and $T_f>T$ due to continuous flow of energy from the effective fractal fluid (chosen as DE) to the normal fluid (considered as DM) and consequently, the thermodynamical equilibrium is violated. Now, from thermodynamical point of view, one may consider the equilibrium temperature $T_0$ as the (modified) Hawking temperature \cite{Chakraborty:2012cw} 
 $$i.e., ~T_0=\frac{H^2 R_h}{2 \pi}|_{a=a_0},$$ 
 where $R_h$ is the geometric radius of the horizon, bounding the Universe.\\
Now, one can investigate the entropy variation of both the fluids under consideration as well as the total entropy variation of the isolated thermodynamical system. If $S_n$ and $S_f$ be the entropies of the normal fluid and effective fractal fluid then using Clausius relation to the individual fluids one obtains
 $$T \frac{dS_n}{dt}=\frac{dQ}{dt}=\frac{dE}{dt}+p\frac{dV}{dt},$$
 and
 $$T_f \frac{dS_f}{dt}=\frac{dQ_f}{dt}=\frac{dE_f}{dt}+p_f\frac{dV}{dt},$$
 where $(Q,~E=\rho V)$ and $(Q_f,~E_f=\rho_f V)$ are the corresponding amount of heat and energy of the two fluid components with $V=\frac{4}{3} \pi R_h^3$ being the volume bounded by the horizon. Assuming the whole Universe bounded by the horizon to be isolated in nature, the heat flow $(Q_h)$ across the horizon is balanced by the heat flow through the two fluid components i.e.,
 $$Q_h=-(Q+Q_f).$$
 Moreover, from the point of view of non--equilibrium thermodynamics, it is desirable to consider the contributions from irreversible fluxes of energy transfer to the total entropy variation as \cite{Saha:2014uwa}
 $$\frac{dS_T}{dt}=\frac{dS_n}{dt}+\frac{dS_f}{dt}+\frac{dS_h}{dt}-A_f \dot{Q_f} \ddot{Q_f}-A_h \dot{Q_h} \ddot{Q_h},$$
 where $S_h$ stands for the entropy of the horizon, $A_f$ and $A_h$ are the constants in the energy transfer between the fluid components within the horizon and across the horizon respectively. Normally, for FLRW model, trapping (i.e., apparent) horizon is chosen as the boundary of the Universe. However, in the perspective of the present accelerating phase of the Universe one may consider event horizon as the bounding horizon.
 \section{Evolution of fractal Universe and deceleration parameter}
 At first, the variation of the deceleration parameter will be examined in fractal cosmology. By introducing $\Omega=\frac{\rho}{3H^2}$ and $\Omega_f=\frac{\rho_f}{3H^2},$ the density parameters for the normal fluid and the effective fractal fluid one gets
 
  \begin{equation} 
  \Omega+\Omega_f=1,\label{f28}
  \end{equation}
 from the Friedmann equation (\ref{f4.1}). Then the deceleration parameter $q$ takes the form (\cite{span})
 
  \begin{equation} 
 q=\frac{1}{2}+\frac{3}{2}(\Omega \gamma+\Omega_f \gamma_f).\label{f29}
  \end{equation}
  
  Now solving equations (\ref{f28}) and (\ref{f29}) for $\Omega$ and $\Omega_f$ one obtains
  
  \begin{equation} 
\Omega=\frac{\left[2q-(1+3\gamma_f)\right]}{3(\gamma-\gamma_f)},\label{f30}
  \end{equation}
  and 
   \begin{equation} 
   \Omega_f=\frac{\left[(1+\gamma)-2q\right]}{3(\gamma-\gamma_f)}.\label{f31}
   \end{equation}
   
   Due to non--negativity of the density parameters, the deceleration parameter has a lower bound and an upper bound i.e.,
   
    \begin{equation} 
    \frac{1+3\gamma_f}{2}\leq q \leq \frac{(1+\gamma)}{2}.\label{f32}
    \end{equation}
    
 During the dominance of the effective fractal fluid $(i.e., \frac{1}{2} \leq \Omega_f \leq 1 ~\mbox{and} ~0 \leq \Omega \leq \frac{1}{2})$ the above inequality (\ref{f32}) modifies to 
 
 \begin{equation} 
 \frac{1+3\gamma_f}{2}\leq q \leq  \frac{1}{2}+\frac{3}{4}(\gamma+\gamma_f).\label{f33 }
 \end{equation}
 
 In particular, if $\gamma_f=-1$ ($i.e.,$ on the phantom barrier) then one has
 $$-1 \leq q \leq -\frac{1}{4}.$$
 
 Further, if one consider the above two fluid system as effective single fluid system then effective equation of state parameter for the single fluid is given by 
 $$\gamma_e=\Omega \gamma+\Omega_f \gamma_f=\gamma_f+\Omega(\gamma-\gamma_f),$$
 and as expected it is independent of the interaction term $Q$. Also the expression for the deceleration parameter takes the form 
 
 \begin{equation} 
 q=\frac{1}{2}(1+3\gamma_e),\label{f34}
 \end{equation}
 the usual definition of the deceleration parameter for a single fluid.\\
 
 Suppose $t_c$ be the time instant at which both the fluids have identical energy densities i.e.,
 
 \begin{equation} 
\rho(t_c)=\rho_f(t_c)=\rho_c~\mbox{(say)},\label{f35}
 \end{equation}
 and $a_c=a(t_c)$ be the value of the scale factor at this instant. Suppose one introduces a dimensionless variable (\cite{span}, \cite{jponce})
 \begin{equation} 
u=\frac{a}{a_c}.\label{f36}
 \end{equation}
\begin{figure}
	\centering
	\includegraphics[width=0.45\textwidth]{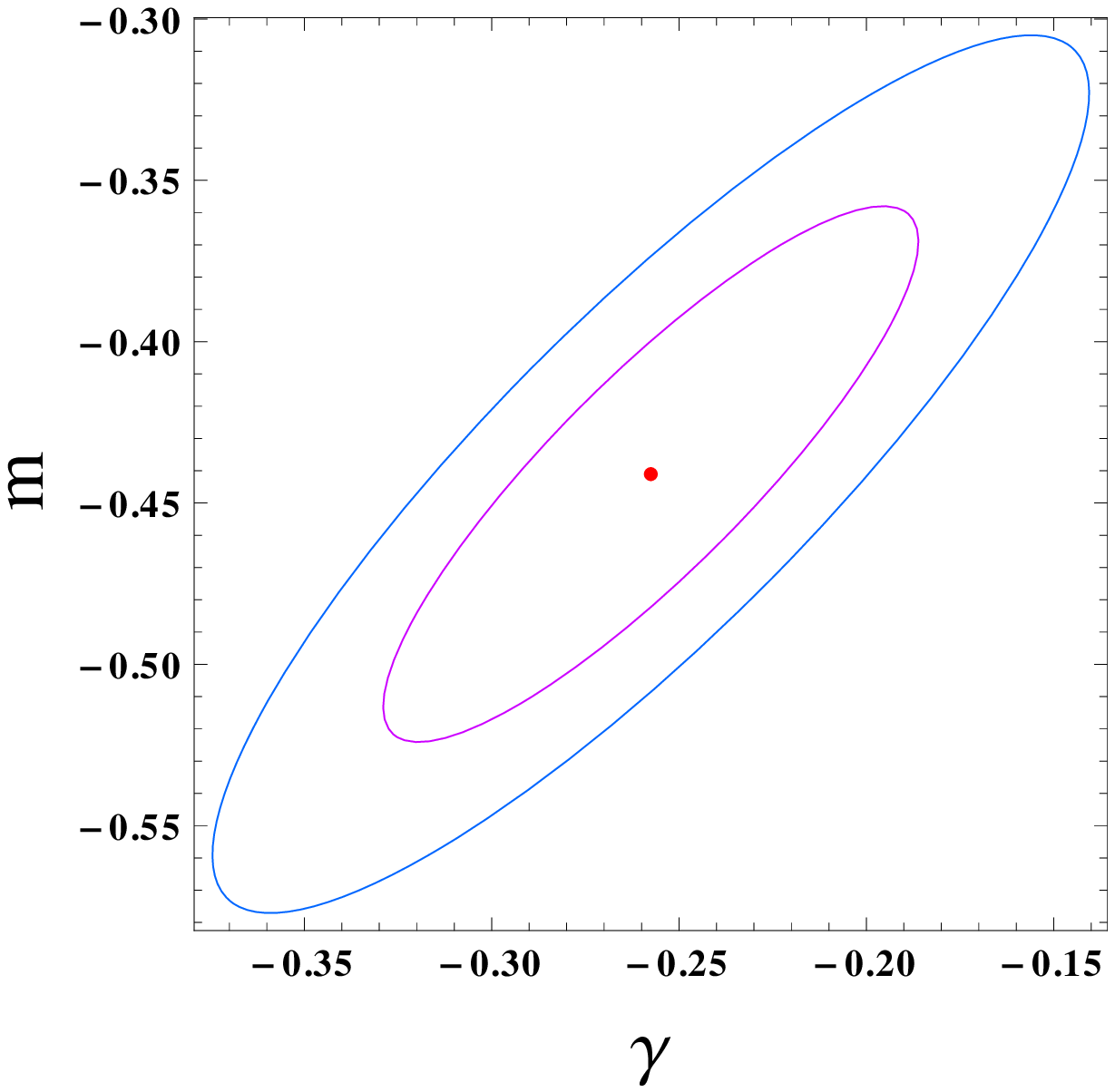}\vspace{0.5cm}\includegraphics[width=0.45\textwidth]{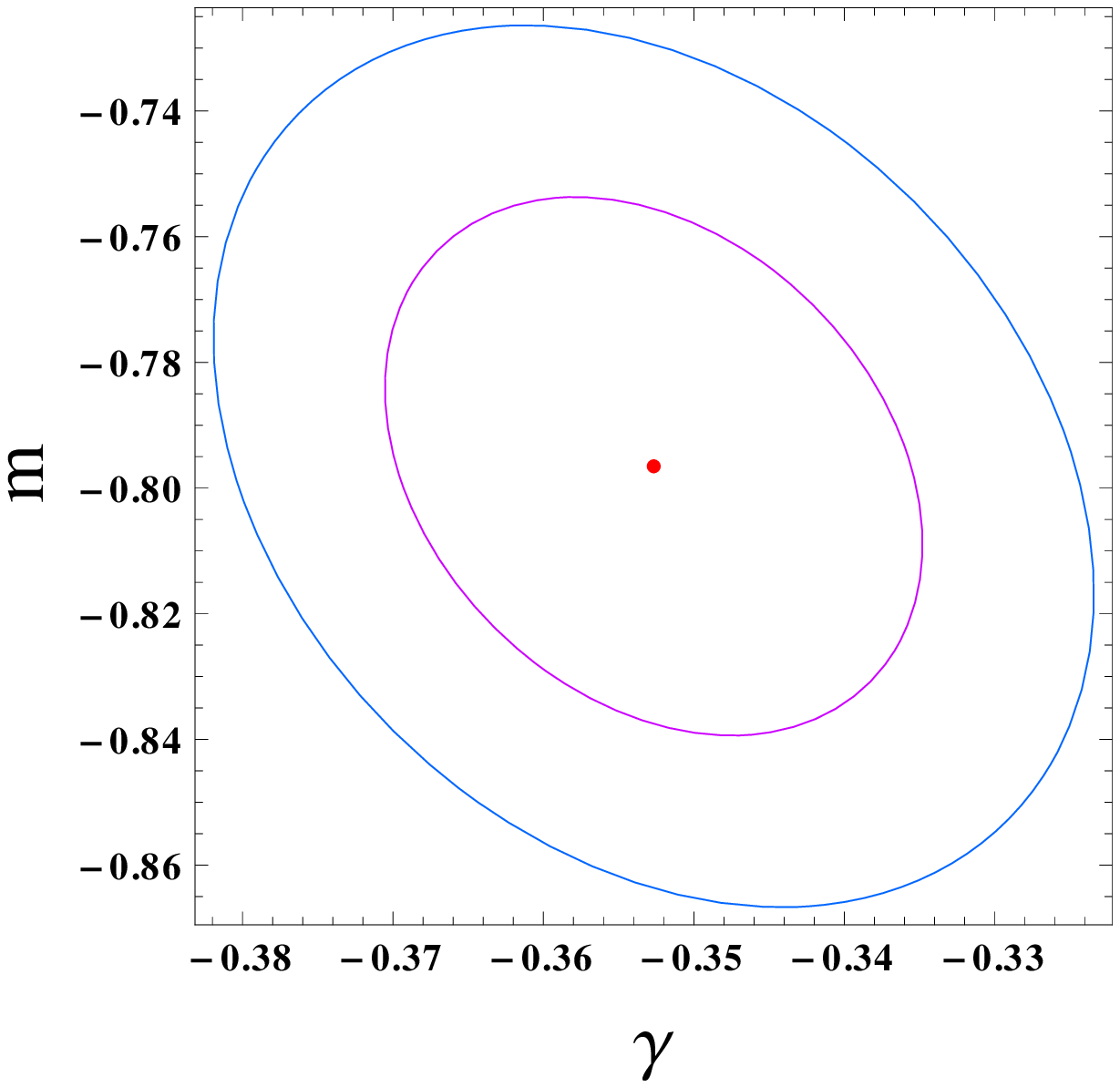}\\
	\caption{The figure shows the $1\sigma$ and $2\sigma$ confidence contours for the choice of $v=v_0a^m$ using SNIa+Hubble data sets. The left and right panels represent the $\gamma-m$ model parameter spaces for the choices $v_0=3.5$ and $w=-0.5$,~$w=0$ respectively. In each panel, the large dot represents the best-fit values of the model parameters.  }
	\label{fig1}
\end{figure}
 \begin{figure}
 	\centering
 	\includegraphics[width=0.5\textwidth]{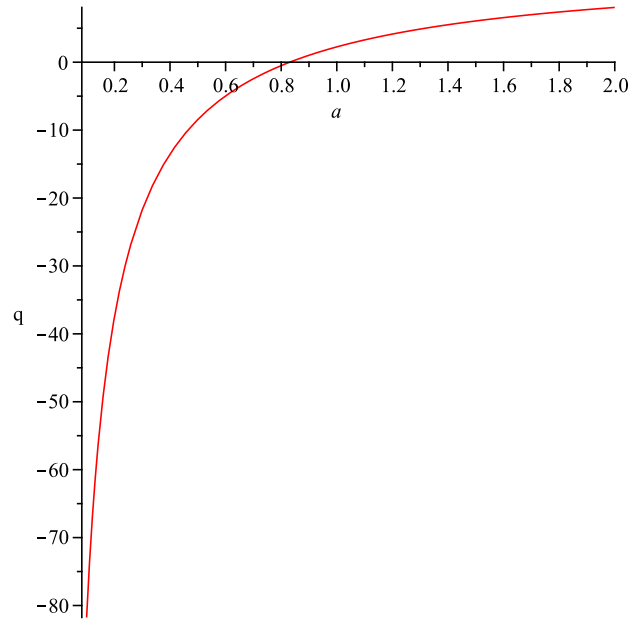}\\
 	\caption{The figure shows the evolution of $q$ using the best-fit values of the parameters $\gamma$ and $m$ arising from the combined analysis of SNIa+Hubble data set for $w=-0.5$ from figure \ref{fig1}}
 	\label{fig2}
 \end{figure}
 and then integrating the continuity equations (\ref{f10}) and (\ref{f11}) using the equations (\ref{f12}) and (\ref{f13}) one has the expressions for energy densities as 
 
  \begin{equation} 
  \rho=\frac{\rho_c}{u^3} \mu(u)~~\mbox{and}~~\rho_f=\frac{\rho_c}{u^3} \gamma(u),\label{f37}
  \end{equation}
  with
   \begin{equation} 
  \mu(u)=exp\left[-3\int_1^u \frac{\gamma^{(eff)} (x)}{x} dx \right]~~\mbox{and}~~\gamma(u)=exp\left[-3\int_1^u \frac{\gamma_f^{(eff)} (x)}{x} dx \right]. \label{f38}
   \end{equation}
   
   As a result the density parameters take the forms 
   
    \begin{equation} 
    \Omega=\frac{\mu(u)}{\mu +\gamma}~~\mbox{and}~~\Omega_f=\frac{\gamma(\mu)}{\mu+\gamma},\label{f39}
    \end{equation}
   and hence the expression for the deceleration parameter becomes
 
  \begin{equation} 
  q(u)=\frac{1}{2}+\frac{3}{2}\left[\frac{\mu \gamma+\gamma \gamma_f}{\mu+\gamma}\right].\label{f40}
  \end{equation}
  
  Now differentiation of the above equation with respect to $u$ with the help of the equations (\ref{f38}) for $\mu$ and $\gamma$ one obtains  (after a bit simplification)
  
   \begin{equation} 
   \frac{dq}{du}=\frac{3}{2}\left( \Omega \frac{d\gamma}{du}+ \Omega_f \frac{d\gamma_f}{du}\right)+\frac{9}{2u} \Omega \Omega_f \left[-(\gamma-\gamma_f)^2-\frac{\dot{v}(1+\gamma)(1+r)}{3vH} (\gamma-\gamma_f)\right].\label{f41}
   \end{equation}

   Normally, the equation of state parameter $\gamma$ is positive and decreases with the evolution (or remains constant) $i.e., \frac{d\gamma}{du} \leq 0$ while $\frac{d\gamma_f}{du}$ may have any sign. Thus, $\frac{dq}{du}$ may change sign more than once during the evolution $i.e, q(u)=0$ may have more than one real solution. Hence Universe may have more than one transition from deceleration to acceleration and vice versa during the process of evaluation of the Universe. However, it should be noted that if $\dot{v}=0$ $i.e., v$ is a constant then fractal effects will be eliminated and one has $\frac{dq}{du} \leq 0 ~i.e.,$ $q$ is a decreasing function of $u$ and $q(u)=0$ has only one real solution. Therefore, one may conclude that fractal effect has a significant impact in the transition from decelerating phase to accelerating phase or in the reverse way.

   \section{Present fractal model and the observed data:}
   
   In fractal cosmology the cosmic fluid in the form of perfect fluid with constant equation of state: $p=\gamma \rho $, ($\gamma$ a constant) is chosen. Regarding the choice of the fractal function $v$ (i.e., measure weight) the monomial \cite{gcal} of the form 
   	  \begin{equation}
   	  v(t)=v_0 t^{-\beta},\label{pf1.1}
   	  \end{equation}
 is chosen with $\beta=4(1-\alpha)$. Here $\alpha>0$, is related to the Hausdorff dimension of the physical space-time. At short scale UV regime if the inhomogeneities are small then the modified Friedmann equations (\ref{f4}) and (\ref{f5}) can be interpreted as background for perturbations rather than a self consistent dynamics. Now eliminating $\rho$ between (\ref{f4}) and (\ref{f5}) (using the equation of state $p=\gamma \rho$) the gravitational constraint equation so obtained is a Riccati equation in $H$, which gives the background expansion without knowledge of the matter content. Hence this over determination of the dynamics rules out the vacuum solution (i.e., $\rho=0=p$). So the above measure may be justified at early times. On the otherhand, if the choice of the measure weight is chosen as a monomial of the scale factor i.e.,
 \begin{equation}
 v=v_0a^{m},~~m<0,\label{pf1.2}
 \end{equation} 
 then this choice of $v$ is similar to the above choice (i.e., in equation (\ref{pf1.1})) in the matter dominated era as scale factor grows as power-law form. So the parameter `$m$' is related to the Hausdorff dimension of the physical space-time. Then the cosmic evolution of the fractal Universe for the above functional form of the fractal function `$v$' has been studied as follows:.  \\

   	 The above monomial form of the fractal function can be used in the energy conservation equation (\ref{f8})  to obtain:~~~~
     
    \begin{equation}
      \rho=\rho_0a^{-\mu}~,~ \mu=(1+\gamma)(m+3),\label{pf1}
    \end{equation}
       and consequently the Hubble parameter can be expressed in terms of the scale factor as
    \begin{eqnarray} 
    H^2&=&\frac{2\rho_0 a^{-\mu} }{6(1+m)- w m^2v_0^2a^{2m}},\nonumber\\
    &=&\frac{H_0^2 a^{-\mu}\big(6(1+m)-wm^2v_0^2\big)}{\big(6(1+m)-wm^2v_0^2a^{2m}\big)}\label{pf2}
    \end{eqnarray}
    
    where $H_0$ is the present value of $H(z)$.\\
    In this case, the deceleration parameter takes the form
    \begin{equation}
    q=-1+\frac{1}{2+m}\left(\frac{w m^2v_0^2(1-\gamma)a^{2m}}{2}+ 3(1+\gamma)(1+m)-m+m^2\right).\label{pf3}
    \end{equation}

   Finally, the present fractal model has also been compared with the observed data, namely Type Ia Supernova (SNIa) constraints and data of the observational Hubble parameter. The total $\chi^2$ function is defined as,
    \begin{equation}
    \chi^2_t=\chi^2_{SNIa}+\chi^2_H,\label{eqchi}
    \end{equation}
    where the individual $\chi^2$ for each data set is calculated as follows:\\
    
    {$\bullet$ {\bf $ \chi^2$ for SNIa data}}\\
    The present phase of accelerated expansion was first pointed out by the data from SNIa observations. In the SNIa observations, the luminosity distance plays an important role and it is given by 
     \begin{equation}
    d_L(z)=\frac{(1+z)}{H_0}\int^z_0 \frac{v(z')~dz'}{h(z')},~~h(z)=\frac{H(z)}{H_0}.\label{eq50}
    \end{equation}
    
    By definition, the distance modulus $\mu$ for any SNIa at a redshift $z$, is written as 
    $$\mu(z)=5\log_{10} \big(\frac{H_0d_{L}(z)}{1Mpc}\big)+\mu_0$$
   where, $\mu_0$ is a nuisance parameter that should be marginalized. Thus the expression for $ \chi^2$ has the form:
   $$ \chi^2_{SNIa}(\mu_0, \theta)=\sum^{580}_{i=1}\frac{\big[\mu^{th}(z_i, \mu_0, \theta)-\mu^{obs}(z_i)\big]^2}{\sigma^2_{\mu}(z_i)} ,$$
  where, $\mu^{th}$ and $\mu^{obs}$ denote the theoretical distance modulus and observed distance modulus respectively. Also, $\theta$ denotes any model parameter and $ \sigma_{\mu}$ denotes the uncertainty in the distance modulus measurements for each data point. In this work, we have used the Union2.1 compilation data \cite{Suzuki:2011hu} containing 580 data points for $\mu$ at different redshifts.\\
  
  Following ref \cite{Lazkoz:2005sp} and marginalizing $\mu_0$, one can write
  $$\chi^2_{SNIa}(\theta)=X(\theta)-\frac{Y
  	(\theta)^2}{Z(\theta)},$$
  where,
  $$X(\theta)=\sum^{580}_{i=1}\frac{\big[\mu^{th}(z_i, \mu_0=0, \theta)-\mu^{obs}(z_i)\big]^2}{\sigma^2_{\mu}(z_i)},$$
   
   $$Y(\theta)=\sum^{580}_{i=1}\frac{\big[\mu^{th}(z_i, \mu_0=0, \theta)-\mu^{obs}(z_i)\big]}{\sigma^2_{\mu}(z_i)},$$
   and
    $$Z(\theta)=\sum^{580}_{i=1}\frac{1}{\sigma^2_{\mu}(z_i)}.$$
    
    {$\bullet$ {\bf $ \chi^2$ for Hubble data}}\\
    Measurements of Hubble parameter are also useful to constraint the parameters in the cosmological models. The relevant $\chi^2$ for the normalized Hubble parameter data set is given by 
       $$\chi^2_H (\theta)= \sum^{30}_{i=1}\frac{\big[h^{th}(z_i, \theta)-h^{obs}(z_i)\big]^2}{\sigma^2_{h}(z_i)}.$$
       
       The error in $h(z)$ is given by \cite{Mamon:2015osa}
       
       $$\sigma_h=h\sqrt{\big(\frac{\sigma_H}{H}\big)^2+\big(\frac{\sigma_{H_0}}{H_0}\big)^2}. $$
       
       Here, one has  used the 30 data points of $H(z)$ measurements \cite{Moresco:2012jh}, \cite{Moresco:2015cya}, \cite{Simon:2004tf}, \cite{Stern:2009ep}, \cite{Zhang:2012mp}, \cite{Samushia:2012iq}, \cite{Delubac:2014aqe}, \cite{Ding:2015vpa}. The present value of $H(z)$ is taken from ref \cite{Riess:2016jrr}.\\
       
       For the (SNIa+Hubble) data set, one can obtain the best--fit values of parameters by minimizing the $\chi^2$ function, as given in equation (\ref{eqchi}). In this work, the 1$\sigma$ and 2$\sigma$ confidence contours on $\gamma-m$ model parameter spaces are plotted using the combined SNIa+Hubble data set, as shown in figure \ref{fig1}. It deserves to mention here that the $\chi^2$ analysis has been done by fixing $v_0=3.5$  and $w=-0.5$,~$w=0$ respectively. The best-fit values are obtained as $\gamma=-0.257$ and $m=-0.441$ ($\chi^2=582.55$) for $w=-0.5$, while it is obtained as $\gamma=-0.352$ and $m=-0.796$ ($\chi^2=598.93$) for $w=0$. Finally, using these best-fit values for $\gamma$ and $m$ the evolution of the deceleration parameter has been shown graphically in figure \ref{fig2} with $w=-0.5$ only. 
    
 \section{Results and Discussions} 
       The present work is an attempt of examining the fractal cosmological model from the point of view of recent observations. Rather than introducing any dark fluid the fractal FRW model is chosen for the description of the  Universe. The modified Friedmann equations are shown to be equivalent to the usual Friedmann equations for interacting two fluid system--of which one is the usual fluid and the other is termed as effective fractal fluid. Also it has been shown that the present model  can be considered as a particle creation mechanism in which the normal matter particles are created while the effective fractal particles are annihilated in course of evolution . 
       
      From thermodynamical consideration the temperature of the individual fluids are evaluated and the entropy variation is determined both for the individual fluids as well as for the total entropy of the system bounded by the horizon.
      
      The fractal function is chosen as a power law form (i.e., monomial) of the scale factor and the deceleration parameter is evaluated. It is found from figure \ref{fig2} that the deceleration parameter has one transition for the above monomial choice of the fractal function. From the graph one may conclude that the present cosmological model with monomial form of the fractal function describes the evolution of the Universe from early inflationary era to the matter dominated phase for the choice $w=-0.5$. It is to be noted that for this graph of deceleration ($q$) parameter, one uses the best fit value of the parameters. For $w=0$ , $q$ is a negative constant, so this choice of $w$ describes only a constant accelerating era. Thus, the fractal cosmological model can describe the evolution from the early inflationary era to the matter dominated epoch. From the comparison with the observation (in figure \ref{fig1}) it is found that the parameter $m$ is always negative ($-0.52<m<-0.36$) with in $2\sigma$ confidence level and it is consistent with the second law of thermodynamics. Therefore from the above analysis one may conclude that the monomial form of the fractal function is consistent with recent observations particularly for describing the early and intermediate era of evolution.
  
 \section*{Acknowledgments} 
 Author DD thanks Department of Science and Technology
 (DST), Govt. of India for awarding Inspire research fellowship (File No: IF160067). SD thanks to the financial support from Science and Engineering Research Board (SERB),
 Govt. of India, through National Post-Doctoral Fellowship Scheme with
 File No: PDF/2016/001435. Also, SD thankful to the Department
 of Mathematics, Jadavpur University where a part
 of the work was completed. SC thanks Science and Engineering Research Board (SERB) for awarding MATRICS Research Grant support (File No: MTR/2017/000407) and Inter University Center for Astronomy and Astrophysics (IUCAA), Pune, India for their
 warm hospitality as a part of the work was done during a visit.

\end{document}